\documentstyle[epsfig]{mn}
\begin{document}

\title[XTE~J1118+480]
{
Spectral evidence for a powerful compact jet from XTE~J1118+480
}
\author[R.~P.~Fender et al.]
{R. P. Fender$^1$, R. M. Hjellming$^2$,  
R. P. J. Tilanus$^3$,
G. G. Pooley$^4$, J. R. Deane$^5$,\cr
R. N. Ogley$^6$, 
R. E. Spencer$^7$
\\
$^1$ Astronomical Institute `Anton Pannekoek', University of Amsterdam,
and Center for High Energy Astrophysics, Kruislaan 403, \\
1098 SJ, Amsterdam, The Netherlands {\bf rpf@astro.uva.nl}\\
$^2$ National Radio Astronomy Observatory, Socorro, NM 87801, USA\\
$^3$ Joint Astronomy Centre, 660 N. A'ohoku Pl., Hilo, Hawaii, USA\\
$^4$ Mullard Radio Astronomy Observatory, Cavendish Laboratory,
Madingley Road, Cambridge CB3 OHE {\bf ggp1@cam.ac.uk}\\
$^5$ Institute for Astronomy, University of Hawaii, 2680 Woodlawn
Drive, Honolulu, HI 96822, USA\\
$^6$ Service d'Astrophysique, CEA Saclay, F-91191 Gif sur Yvette,
Cedex, France\\
$^7$ University of Manchester, Nuffield Radio Astronomy Laboratories,
Jodrell Bank, Cheshire, SK11 9DL\\
}

\maketitle

\begin{abstract}

We present observations of the X-ray transient XTE J1118+480 during
its Low/Hard X-ray state outburst in 2000, at radio and sub-millimetre
wavelengths with the VLA, Ryle Telescope, MERLIN and JCMT.  The
high-resolution MERLIN observations reveal all the radio emission (at
5 GHz) to come from a compact core with physical dimensions smaller
than $65$ d(kpc) AU.  The combined radio data reveal a persistent and
inverted radio spectrum, with spectral index $\sim +0.5$. The source
is also detected at 350 GHz, on an extrapolation of the radio
spectrum. Flat or inverted radio spectra are now known to be typical
of the Low/Hard X-ray state, and are believed to arise in synchrotron
emission from a partially self-absorbed jet. Comparison of the radio
and sub-millimetre data with reported near-infrared observations
suggest that the synchrotron emission from the jet extends to the
near-infrared, or possibly even optical regimes. In this case the
ratio of jet power to total X-ray luminosity is likely to be $P_{\rm
J} / L_{\rm X} >> 0.01$, depending on the radiative efficiency and
relativistic Doppler factor of the jet. Based on these arguments we
conclude that during the period of our observations XTE J1118+480 was
producing a powerful outflow which extracted a large fraction of the
total accretion power.

\end{abstract}



\section{Introduction}

XTE J1118+480 is a new transient X-ray source, discovered in soft
($\leq 12$ keV) X-rays by the all-sky monitor (ASM) onboard the Rossi
X-ray Timing Explorer (RXTE) in 2000 March (Remillard et
al. 2000). Within days optical (Uemura, Kato \& Yamaoka 2000a), radio
(Pooley \& Waldram 2000) and hard X-ray (Wilson \& McCollough 2000)
counterparts to the new transient were established.  Uemura et
al. (2000b) report in detail the discovery and subsequent photometric
observations of the optical counterpart.  Garcia et al. (2000)
reported strong H$\alpha$ emission, and established that the
interstellar extinction towards the source was very low. This fact
allowed for the first time EUV observations of an X-ray transient
(Hynes et al. 2000). Patterson (2000) further reported rapid and
erratic variability in the optical flux from the system. Merloni, Di
Matteo \& Fabian (2000) have modelled the rapid optical variability as
synchrotron emission from magnetic flares in a corona above the
accretion disc.

The X-ray properties of XTE J1118+480, both in timing and spectra,
indicate that the system is a black hole (candidate) in the Low/Hard
X-ray state (Revnivtsev, Sunyaev \& Borozdin 2000; Wood et al. 2000),
characterised by strong low-frequency variability and a hard power-law
component extending to $\ga 100$ keV which dominates any contribution
in X-rays from an accretion disc. The radio spectrum of the system is
persistently inverted in the range 1--22 GHz (Dhawan et al. 2000;
Hynes et al. 2000) with a spectral index ($\alpha$, where $S_{\nu}
\propto \nu^{\alpha}$) of about +0.5.
Such flat or inverted radio spectra have
been established as a ubiquitous observational characteristic of the
Low/Hard X-ray state in both peristent and transient black hole X-ray
binaries (Fender 2000a,b).

\section{Observations}

During a 100-day period between MJD 51620 -- 51720 XTE
J1118+480 exhibited a steady level of X-ray and radio emission. The
observations reported below were obtained during this steady
phase. More recently the source has declined rapidly again (since
around MJD 51745).

\subsection{VLA}

Regular monitoring of XTE J1118+480 has been undertaken by the Very
Large Array (VLA) throughout the 2000 outburst. During the interval
MJD 51620 -- 51720, the mean flux densities at 1.4, 8.4 and 22.5 GHz
were $2.6 \pm 0.4$, $6.5 \pm 0.7$ and $9.3 \pm 1.2$ mJy respectively,
where uncertainties reflect the statistical deviation or the sample
(intrinsic source variability and/or calibration errors) rather than
the significance of the detections.

\subsection{RT}

The Ryle Telescope (RT) was used to monitor the flux density at 15.2 GHz,
using techniques similar to those described in Pooley \& Fender (1997). 
The phase calibrator used was J1110+440, and the flux-density scale 
established by observations of 3C48 and 3C286. The mean flux density
measured by the RT over the period under discussion was $9.0 \pm 1.0$ mJy.

\subsection{MERLIN}

\begin{figure}
\leavevmode\epsfig{file=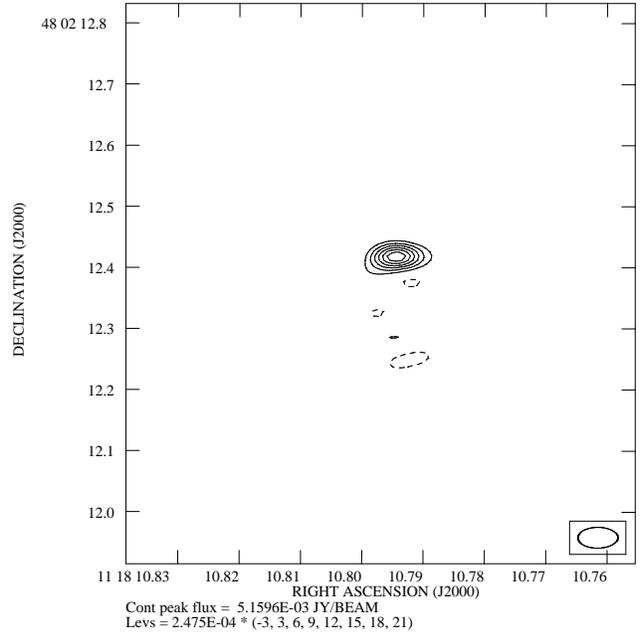, angle=0, width=8.5cm}
\caption{5 GHz MERLIN image of XTE J1118+480, revealing a compact
unresolved core. Comparison with RT and VLA monitoring indicates that
all the radio emission arises within this core. The synthesised beam,
$65 \times 35$ mas, is indicated in the lower right-hand corner.
}
\end{figure}

XTE J1118+480 was observed by MERLIN on 2000 Apr 5 1650--2053 UT, in
standard continuum observing mode at 4.994 GHz with a 15 MHz
bandwidth, and primary and secondary flux calibrators 3C286 and
OQ208.  To phase-calibrate the target source a 6 minute cycle nodding
between the target and phase calibrator 1107+485 was used.  Even with
the short integration time and therefore incomplete $uv$ coverage, a
fully calibrated image of 1118+480 was obtained (Fig 1). 
The image has a 1
$\sigma$ noise level of 248 $\mu$Jy/beam with a synthesised beam of 65
$\times$ 35 mas at 90 degrees from north.
XTE J1118+480 is an unresolved point source at co-ordinates RA(2000)
11 18 10.794 Dec(2000) +48 02 12.42 with a flux density of 5.16
mJy. The MERLIN run was simultaneous with an observation with
the RT, which measured a flux density of $8.7 \pm 0.3$ mJy at 15 GHz,
confirming the inverted spectrum ($\alpha = +0.5$).

\subsection{JCMT}

XTE J1118+480 was observed at 350 GHz (850 $\mu$m) with the SCUBA
instrument on the James Clerk Maxwell Telescope (JCMT) on 2000 May 30
and 31. The atmospheric optical depth at 850 $\mu$m was variable but
around 0.4 during both observations (based on skydips and measurements
from the nearby Caltech Submillimetre Observatory). Flux calibration
was performed using Uranus and CIT6. XTE J1118+480 is clearly detected
on both nights (with no variability at $>20$\% level between the two
runs). The mean flux density from the two nights' observations is $41
\pm 4$ mJy, the error being dominated by the uncertainty in the flux
conversion factor.

Another observation of the source at 350 GHz was performed on 2000
September 9, at which time only a (3$\sigma$) upper limit of 21 mJy
was obtained. Given the decay in radio emission from XTE J1118+480 by
this time this supports the association of the submillimetre source
with the X-ray transient.

\section{Discussion}

\begin{figure*}
\leavevmode\epsfig{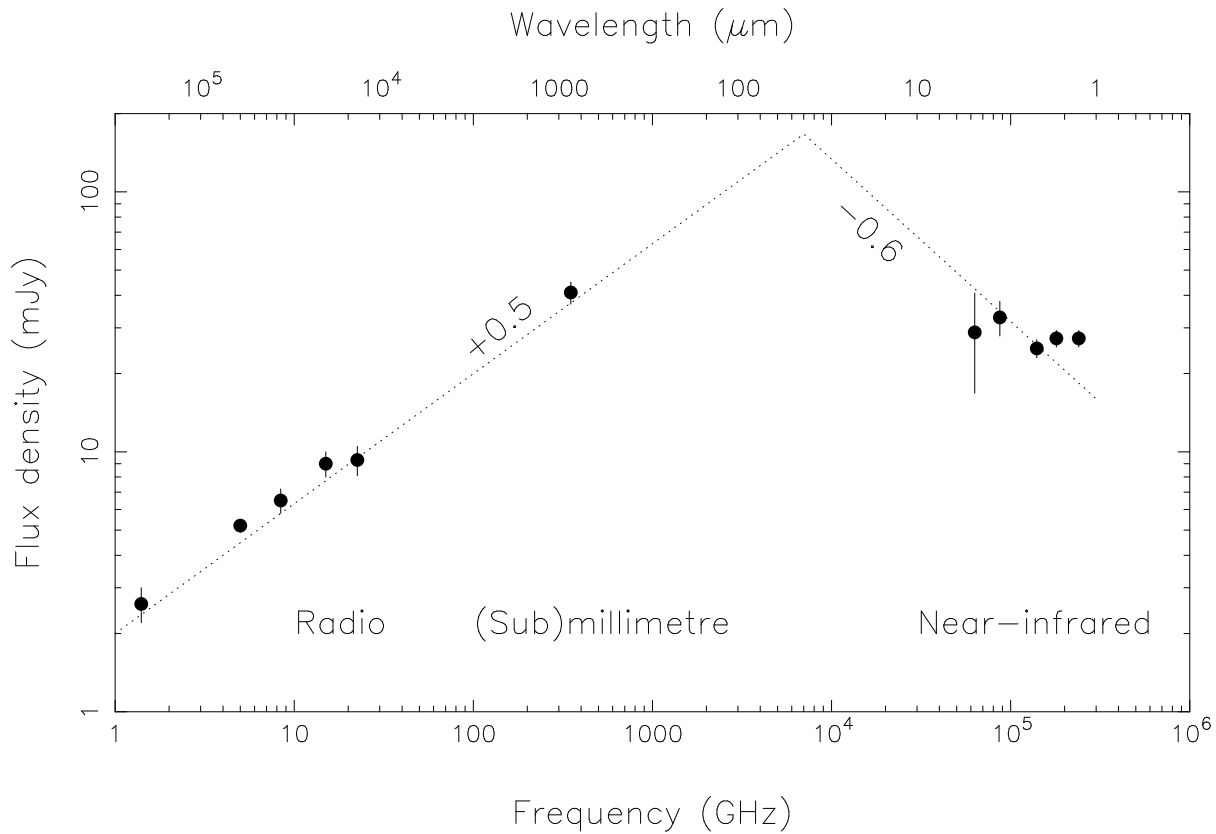}
\caption{Radio -- Sub-mm -- Near-infrared broadband spectrum
of XTE J1118+480 during the extended, steady, period in the Low/Hard
X-ray state (MJD 51620-51720). Note the inverted radio spectrum which
extrapolates directly to the sub-mm measurement at 350 GHz; this
component is likely to be synchrotron emission from a compact,
partially self-absorbed jet. The near-infrared data are from Hynes et
al. (2000); there is strong evidence for an additional contribution at
longer wavelengths (see text and Fig 3), which may be the optically
thin extension of the jet spectrum. Also indicated on the figure is a
double power-law corresponding simplistically to self-absorbed and
optically thin regimes in the jet. }
\end{figure*}

In Fender (2000a,b) it has been shown that all X-ray binaries in the
Low/Hard X-ray state produce a flat or inverted radio spectrum
(i.e. $\alpha \geq 0$), which is likely to extend spectrally to the
millimetre or infrared regimes. This spectral component is likely to
arise in a partially self-absorbed synchrotron emission from a
relativistic outflow or jet from each system. The power into this
outflow appears to be a significant ($\geq 5$\%) and approximately
fixed fraction of the accretion luminosity. In the following we
discuss the observations of XTE J1118+480 in the context of such a
model.

\subsection{Compact core}

The flux density of the unresolved component imaged with MERLIN is
consistent with the radio spectrum measured by the VLA and RT,
therefore we can be confident that all the radio emission from XTE
J1118+480 arises within a region smaller than the MERLIN beam.  We
note that the (one-sided) radio jet from Cyg X-1 in the Low/Hard state
has an angular extent of $\sim 15$ mas at 8 GHz (Stirling, Spencer \&
Garrett 1998; Stirling et al. 2000), at a distance of $\sim 2$
kpc. Given a comparable GHz flux density we might expect a similar
angular extent from XTE J1118+480, so the 35 $\times$ 65 mas beam
probably failed to resolve the core by less than a factor of 10 (and
the system may have been resolveable with the VLBA, depending on
orientation).

\subsection{Broadband spectrum}

Fig 2 plots the broadband spectrum (SED) of XTE J1118+480,
from radio to near-infrared wavelengths, during the $\sim 100$-day
steady period in the Low/Hard X-ray state (see also Hynes et al. 2000
for a more extensive SED, but without the sub-millimetre
datum). Within uncertainties dominated by the non-simultaneity of the
observations, the radio data correspond to a power-law of spectral
index $\sim +0.5$. The JCMT measurement sits exactly on an
extrapolation of this power-law to the sub-millimetre regime. We
therefore consider it most likely that emission at 350 GHz is an
extension of the inverted spectral component from the radio.
Furthermore, the significant decrease in the 350 GHz flux density
when the source was reobserved in 2000 September, by which point
the radio flux density had also dropped dramatically, supports
this interpretation.

It is unlikely that our detection of the system at 350 GHz is exactly
at the high-frequency break of the inverted spectral component,
however comparison with the near-infrared fluxes reported by Hynes et
al. (2000) indicates that some break, probably to an optically thin
spectrum, occurs before $10^{14}$ Hz (Fig 2). Naively assuming that
the inverted spectral component continues with a spectral index of
$+0.5$ and then breaks at one point to an optically thin spectrum with
$\alpha = -0.6$ (a typical value)
which connects with the lowest-frequency near-infrared
point (see below), the break will occur around $\sim 7 \times 10^{12}$
Hz (40 $\mu$m), with a peak flux density of about 150 mJy. This crude
`fit' to the radio--mm--infrared data is plotted in Fig 2.

Three sets of near-infrared observations have been reported, in Hynes
et al. (2000) and Taranova \& Seranova (2000); these are plotted in
Fig 3.  Since the extinction to the source is very low (E(B-V)=0.013
-- Hynes et al. 2000), it is not necessary to deredden the infrared
data.  Bearing in mind that an accretion disc should have a thermal
spectrum with spectral index in the range $1/3 \la \alpha \la 2$, it
appears that there is excess infrared emission at the longest
wavelengths ($\lambda \geq 2$ $\mu$m, ie. in the K-band and
beyond). Hynes et al. (2000) have already noted that another source of
near-infrared flux (beyond an accretion disc) is present.  Since the
sub-mm detection clearly demonstrates that the inverted spectral
component does not cut off in the radio band, it seems plausible to
connect it to the excess emission in the near-infrared.  The reader is
reminded that there is already very strong evidence for synchrotron
emission from the radio through the millimetre to the near-infrared
from the black hole system GRS 1915+105 when in hard X-ray states
(e.g. Fender \& Pooley 1998, 2000 and references therein), and that
Fender (2000b) presents observational evidence that the synchrotron
spectrum may extend to the near-infrared or even optical regimes in
{\em all} black hole X-ray binaries which are in the Low/Hard X-ray state.

\begin{figure}
\leavevmode\epsfig{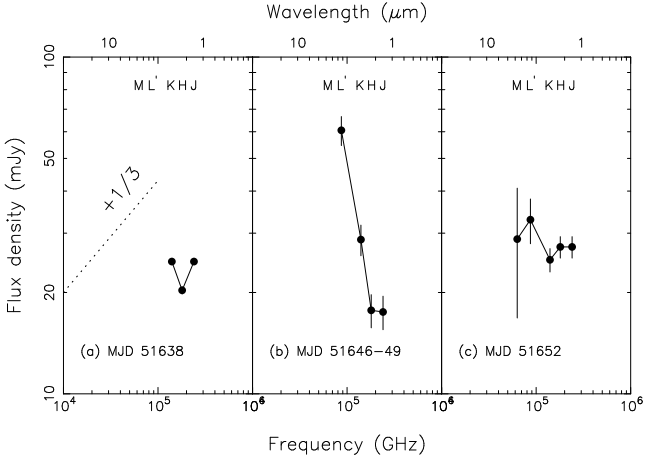}
\caption{Three sets of reported near-infrared observations of XTE
J1118+480, obtained during the period of relatively steady X-ray
state. Thermal emission from e.g. an accretion disc would be expected
to have a positive spectral index (ie. rising with frequency) in the
range 0.3 -- 2.0; in panel (a) a spectral index of +1/3 is indicated.
These data clearly indicate an excess above any such thermal
emission. Data from Hynes et al. (2000; panels (a) and (c)) and
Taranova \& Shenavrin (2000; panel (b)).}
\end{figure}

\subsection{Energetics}

The integrated 1--350 GHz luminosity of the inverted-spectrum
component is $L_{\rm J} = 10^{31} (d/{\rm kpc})^2$ erg s$^{-1}$. When
in the Low/Hard X-ray state, Cyg X-1 has a typical XTE ASM count rate
of 20 ct/sec; it is assumed to lie at a distance of $\sim 2$ kpc and
has an integrated X-ray luminosity (dominated by the power-law
component) of $3 \times 10^{37}$ erg s$^{-1}$ (Di Salvo et
al. 2000). Therefore, under the (reasonable) assumption that all X-ray
binaries in the Low/Hard state have comparable spectra, we scale from
Cyg X-1 to derive the relation $L_{\rm X,LHS} \sim 4 \times 10^{35} R
(d/{\rm kpc})^2$ erg s$^{-1}$, where $R$ is the XTE/ASM 2-12 keV count
rate.  The mean XTE ASM count rate for XTE J1118+480 at the time of
our observations was around 2, so in this case $L_{\rm X} \sim 8
\times 10^{35} (d/{\rm kpc})^2$ erg s$^{-1}$. Broadband X-ray
observations (McClintock et al. 2000) confirm this estimate of the
2--200 keV luminosity to within a factor of two, validating the method
(however those authors also note that the X-ray spectrum may peak at
even higher energies in this source).  Note that simply using measured
soft X-ray fluxes from XTE J1118+480, such as those reported in
Remillard et al. (2000) will underestimate the X-ray luminosity, which
normally peaks around 100 keV in the hard state, but has been
accurately measured for Cyg X-1. From this, $L_{\rm J} / L_{\rm X}
\sim 2 \times 10^{-5}$; this should be considered as a firm lower
limit to the ratio of jet to X-ray luminosities (assuming the
impossible case of a 100\% radiatively efficient jet), unless the 
X-ray spectrum of XTE J1118+480 really turns out to be very different
from that of Cyg X-1 at higher ($> 200$ keV) energies.

If the jet spectrum is something like the dual power-law `fit' shown
in Fig 2, then the integrated luminosity of the {\em inverted}
spectral component is $8 \times 10^{32} (d/{\rm kpc})^2$ erg
s$^{-1}$. However the radiative luminosity is dominated by the extent
of the high-frequency emission; regardless of the spectral form at
lower frequencies, if the jet contributes $\sim 20$ mJy at $3 \times
10^{14}$ Hz then the integrated radiative luminosity $L_{\rm J} \geq
10^{34} (d/{\rm kpc})^2$ erg s$^{-1}$.  ie.  $L_{\rm J} / L_{\rm X}
\geq 0.01$. Even if the jet spectrum peaks at 350 GHz and a power law
(with $\alpha \sim -0.1$) connects directly from the submillimetre to
the infrared regimes $L_{\rm J}$ is reduced from that of the `fit' by
less than 10\%.  The total jet power can be estimated as $P_{\rm J}
\sim L_{\rm J} \eta^{-1} F(\Gamma,i)$, where $\eta$ is the radiative
efficiency of the jet and $F(\Gamma,i) = \Gamma \delta^{-3}$ is a
correction for relativistic motion (Fender 2000b). Assuming $\eta =
0.05$ (which seems reasonable -- Fender 2000b and references therein),
and $F(\Gamma,i)=1$ (ie. no significant relativistic correction), the
jet power will be $\geq 20$\% of the integrated X-ray luminosity, and
therefore a very significant factor indeed for the energetics of the
system. How reasonable is the assumption $F(\Gamma,i) \sim 1$ ? Dubus
et al. (2000) estimate the orbital inclination of the binary to be $30
\leq i \leq 70$ based on optical spectroscopy. For $\Gamma \leq 5$,
$0.2 \leq F(\Gamma,i) \leq 200$; ie. the maximum we could be {\em
overestimating} the jet power by is a factor of 5, whereas we could be
{\em underestimating} it by two orders of magnitude.

\subsection{Optical variability}

Merloni et al. (2000) argue that the rapid optical variability of XTE
J1118+480, and in particular its correspondence with X-ray
variability, indicates that the optical flux is generated in the inner
regions of the accretion disc, and is of nonthermal origin. Rapid
near-infrared variability in phase with X-ray variability has been
directly observed from GRS 1915+105 (Eikenberry et al. 1998; Mirabel
et al. 1998), and in this source there is little doubt that the
infrared flux is associated with powerful relativistic ejections (Fender \&
Pooley 1998, 2000; Dhawan, Mirabel \& Rodriguez 2000). In Cyg X-3
rapid near-infrared flares are also probably associated with a
relativistic jet (Fender et al. 1996).  Since we have argued that the
excess near-infrared flux is associated with the jet emission, and
rapid near-infrared variability seems often to be associated with jet
sources, we suggest that the rapid optical flaring in XTE J1118+480 is
also associated with the jet, probably being optically thin nonthermal
synchrotron emission from the `base' of the jet, very close to the
accretion disc. This is similar to the model of Merloni et al. (2000),
except that we envisage the sites of optical emission as being
associated with a global structure which, at radio wavelengths at
least, is resolved into a collimated outflow.

\section{Conclusions}

We have presented radio and sub-mm observations of the X-ray transient
XTE J1118+840, and combined them with published near-infrared data to
present the broadband spectrum of the system from 1 --
$10^{6}$ GHz during a steady period of Low/Hard X-ray state emission.
The radio spectrum is inverted, with $\alpha = +0.5$, consistent
with flat/inverted spectra from partially self-absorbed jets being a
generic feature of the Low/Hard X-ray state (Fender 2000a,b). Two
lines of argument are presented to argue that this spectral component
extends from the radio regime through to the near-infrared :

\begin{itemize}
\item{The detection of the system at 350 GHz (850 $\mu$m) precisely on
an extrapolation of the inverted radio spectrum. 
This clearly indicates that the inverted spectral
component extends significantly beyond the radio band.}
\item{Excess flux in reported near-infrared measurements. Three sets
of reported near-infrared photometry of XTE J1118+480 all reveal
evidence for excess, apparently non-thermal, infrared emission in
about the K-band and beyond}
\end{itemize}

Using these two observational facts, we argue that the inverted
spectral component extends from the radio regime all the way to the
near-infrared or optical regimes. If this is the case, then the ratio
of jet power to X-ray luminosity is $P_{\rm J} / L_{\rm X} \sim 0.01
\eta^{-1}$, where $\eta$ is the radiative efficiency of the jet (see
Fender 2000b). For discrete ejection events from GRS 1915+105 (which
clearly have a flat/inverted spectrum extending from the radio,
through the mm, to the near-infrared regime) the maximum radiative
efficiency appears to be $\sim 0.05$ (Fender \& Pooley 2000);
therefore the jet power is likely to be a major, if not perhaps
dominant, power output channel for the system in this X-ray state.

It is interesting to think about the size scales associated with the
jet; for a Blandford \& K\"onigl (1979) jet the physical size associated
with emission at a given frequency scales as $\nu^{-1}$. If we assume
the jet to have similar dimensions to that of Cyg X-1, ie. $\sim 30$
AU at 8 GHz, then emission at 350 GHz will arise from a region of size
scale $\la 1$ AU. If the inverted component does peak around $7 \times
10^{12}$ Hz, this would correspond to a size scale of around 0.03 AU,
or $5 \times 10^{11}$ cm. The optically thin emission observed in the
near-infrared will come from even smaller phyiscal scales, whose size
would be best constrained by rapid variability.

XTE J1118+480 is not unique in showing evidence for jet emission
extending to the near-infrared when in the `canonical' Low/Hard X-ray
state. Fender (2000b) argues that the synchrotron spectrum probably
extends to the near-infrared or optical regimes in all X-ray binaries
in this state. Corbel \& Fender (2001) and Corbel et al. (2001)
present further evidence for near-infrared synchrotron emission from
GX 339-4 and XTE J1550-564 when in the Low/Hard X-ray state, a
persistent and transient source respectively. Brocksopp et al. (2000)
finds similarities in the optical properties of Low/Hard state
transients, which may also be related to high-frequency synchrotron
emission associated with the jet. However in most cases the sub-mm and
far-infrared regimes, crucial for establishing the connection between
the radio and near-infared emission, are not explored. Such
observations, admittedly technically difficult, are vital for
confirming our qualitative model and establishing that in the Low/Hard
X-ray state the jet can be a major channel for the output of accretion
power from black holes.  In a more theoretical study Markoff, Falcke
\& Fender (2001) have applied a jet-dominated model to the broadband
radio through X-ray spectrum of XTE J1118+480, and found that even the
power-law X-ray component may arise as in the jet, possibly as direct
optically thin synchrotron emission. In this case more than 90\% of
the power output of the system is in the form of the jet.  While
ironically the model of Markoff et al. (2001) is unable to fit the
JCMT datum, one of the motivations in this paper for arguing for the
presence of a jet, the overall fit to the broadband spectrum is
striking, with important implications for the interpretation of hard
X-ray spectra if correct.

\section*{Acknowledgements}

We would like to thank the staff of the JCMT for approval and
execution of these observations at short notice.  The JCMT is operated
by The Joint Astronomy Centre on behalf of the UK Particle Physics and
Astronomy Research Council (PPARC), the Netherlands Organisation for
Scientific Research and the National Research Council of Canada.
MERLIN is operated as a National Facility by the University of
Manchester at the Nuffield Radio Astronomy Laboratories, Jodrell Bank,
on behalf of the Particle Physics and Astronomy Research Council
(PPARC).  operated by The Observatories on behalf of the PPARC.  We
thank the staff at MRAO for maintenance and operation of the RT, which
is supported by the PPARC.


\begin{thebibliography}{}


\bibitem[]{}
Blandford R., K\"onigl A., 1979, ApJ, 232, 34

\bibitem[]{}
Brocksopp C., Jonker P.G., Fender R.P., Groot P.J., van der Klis M.,
Tingay S.J., 2000, MNRAS, in press, ({\bf astro-ph/0011145})

\bibitem[]{}
Corbel S., Fender R.P., 2001, ApJ, in prep

\bibitem[]{}
Corbel S. et al., 2001, ApJ, submitted

\bibitem[]{}
Dhawan V., Mirabel I.F., Rodriguez L.F., 2000, ApJ, 543, 373

\bibitem[]{}
Dhawan V., Pooley G.G., Ogley R.N., Mirabel I.F., 2000, IAU Circ. 7395

\bibitem[]{}
Di Salvo T., Done C., Zycki P.T., Burderi L., Robba N.R., 2000, ApJ, submitted

\bibitem[]{}
Dubus G., Kim R.S.J., Menou K., Szkody P., Bowen D.V., ApJ, submitted,
({\bf astro-ph/0009148})

\bibitem[]{}
Fender R.P., 2000a, {\em Black hole states and radio jet formation},
L. Kaper, E.P.J. van den Heuvel, P.A. Woudt (Eds), `Black holes in
binaries and galactic nuclei', ESO workshop, Springer-Verlag, in press
{\bf (astro-ph/9911176)}

\bibitem[]{}
Fender R.P., 2000b, MNRAS, in press, ({\bf astro-ph/0008447})

\bibitem[]{}
Fender R.P., Pooley G.G., 1998, MNRAS, 300, 573 

\bibitem[]{}
Fender R.P., Pooley G.G., 2000, MNRAS, 318, L1

\bibitem[]{}
Fender R.P., Bell~Burnell S.J., Williams P.M., Webster A.S., 1996,
MNRAS, 283, 798

\bibitem[]{}
Garcia M., Brown W., Pahre M., McClintock J., Callanan P., Garnavich
P., 2000, IAU Circ. 7392

\bibitem[]{}
Markoff S., Falcke H., Fender R., 2000, ApJ Lett, submitted,
({\bf astro-ph/0010560})

\bibitem[]{}
Hynes R.I., Mauche C.W., Haswell C.A., Shrader C.A., Cui W., Chaty S.,
2000, ApJ,  539, L37

\bibitem[]{} Merloni A., Di Matteo T., Fabian A.C., 2000, MNRAS,
318, L15

\bibitem[]{}
Patterson J., 2000, IAU Circ. 7412

\bibitem[]{}
Pooley G.G., Fender R.P., 1997, MNRAS, 292, 925

\bibitem[]{}
Pooley G.G., Waldram E.M., 2000, IAU Circ. 7390

\bibitem[]{}
Remillard R., Morgan E., Smith D., Smith E., 2000, IAU Circ. 7389

\bibitem[]{} 
Stirling A., Spencer R., Garrett M., 1998, New
Astronomy Reviews, 42, 657

\bibitem[]{} 
Stirling A.M., Spencer R.E., de la Force C.J.,  Garrett M., Fender R.P.,
2000, MNRAS, in prep

\bibitem[]{}
Taranova O., Shenavrin V., 2000, IAU Circ. 7407

\bibitem[]{}
Uemura M., Kato T., Yamaoka H., 2000a, IAU Circ. 7390

\bibitem[]{}
Uemura M. et al., 2000, PASJ, 52, L15

\bibitem[]{}
Wilson C.A., McCollough M.L., 2000, IAU Circ. 7390

\bibitem[]{}
Wood K.S. et al., 2000, ApJ, 544, L45

\end{thebibliography}
\end{document}